\documentclass[universe,review,accept,oneauthor,pdftex,10pt,a4paper]{Definitions/mdpi} 

\newcommand{\msun}{{M}_{\odot}}

\newbox\grsign \setbox\grsign=\hbox{$>$}
\newdimen\grdimen \grdimen=\ht\grsign
\newbox\laxbox \newbox\gaxbox
\setbox\gaxbox=\hbox{\raise.5ex\hbox{$>$}\llap
     {\lower.5ex\hbox{$\sim$}}}\ht1=\grdimen\dp1=0pt
\setbox\laxbox=\hbox{\raise.5ex\hbox{$<$}\llap
     {\lower.5ex\hbox{$\sim$}}}\ht2=\grdimen\dp2=0pt

\newcommand{\lax}{$\mathrel{\copy\laxbox}$}

\firstpage{1} 
\pubvolume{xx}
\issuenum{1}
\articlenumber{5}
\pubyear{2018}
\copyrightyear{2018}
\history{Received: date; Accepted: date; Published: date}
\Title{Approaching the black hole by numerical simulations}

\Author{Christian Fendt*}

\AuthorNames{Christian Fendt}

\address{%
Max Planck Institute for Astronomy, Königstuhl 17, D-69117 Heidelberg, Germany; fendt@mpia.de}
\corres{Correspondence: fendt@mpia.de}

\abstract{
Black holes represent extreme conditions of physical laws.
Being predicted about a century ago, they are now accepted as astrophysical reality by most of the scientific community.
Only recently more direct evidence of their existence has been found - the detection of gravitational waves from black hole
mergers and of the shadow of a supermassive black hole in the center of a galaxy.
Astrophysical black holes are typically embedded in an active environment which is affected by the
strong gravity.
When the environmental material emits radiation, this radiation may carry imprints of the black hole 
that is hosting the radiation source.
In order to understand the physical processes that take place in the close neighbourhood
of astrophysical black holes, numerical methods and simulations play an essential role.
This is simply because the dynamical evolution and the radiative interaction are far too 
complex in order to allow for an analytic solution of the physical equations.
A huge progress has been made over the last decade(s) in the numerical code development, 
as well as in the computer power that is needed to run these codes. 
This review tries to summarize the basic questions and methods that are involved 
in the undertaking of investigating the astrophysics of black holes by numerical means.
It is intended for a non-expert audience interested in an overview over this broad field. 
The review comes along without equations and thus without a detailed expert discussion of the 
underlying physical processes or numerical specifics. 
Instead, it intends to illustrate the richness of the field and to motivate for further reading.
The review puts some emphasis on magnetohydrodynamic simulations, but also touches radiation 
transfer and merger simulations, in particular pointing out differences in these approaches.
}


\begin{document}
%%%%%%%%%%%%%%%%%%%%%%%%%%%%%%%%%%%%%%%%%%

%\setcounter{section}{-1} %% Remove this when starting to work on the template.
%----------------------------------------------------------------------------------------
\section{Introduction}
For many people - the scientists and the public -- black holes are considered as the most 
fascinating objects in the universe.
In principle, black holes simply consist of highly compressed and invisible mass,
solely described by its mass and angular momentum, and are, thus, physically and 
chemically less rich compared to e.g. the Solar atmosphere.
However, the compactness of the accumulated mass distorts the surrounding space time 
in such a way that the physical processes occurring close to a black hole seem to behave 
in contradiction to the every-day experience.

A prime mystery is (and may remain) the interior of a black hole -- nobody knows how the 
innermost parts of a black hole are structured as we do not yet understand the physics at 
work under such extreme conditions. 
Below the black hole {\em horizon} the time and space coordinates change their 
meaning, which may impact our common understanding of causality.
In order to understand the very center - if it exists - the extreme physical conditions 
would ask for a new theory of physics, that is quantum gravity, the combination of general 
relativity and quantum physics.

Astrophysical black holes come in various manifestations - from stellar mass black holes 
that can sometimes be observed as micro-quasars \citep{1999ARA&A..37..409M} to 
{\em supermassive} black holes that reside in the nuclei of active galaxies and weigh 
up to billions of Solar masses \citep{2012ARA&A..50..455F,2013peag.book.....N}.
Intermediate mass black holes weighing some 1000s of Solar masses are discussed but not 
yet confirmed.
Recently, by their emission of gravitational waves, a new flavor of black holes has been 
discovered rather unexpectedly -- black holes with masses of 30-50 Solar masses 
for which the origin is not yet known.

Astrophysical black holes do not exist alone in space, 
but are surrounded by other material that is moving with high speed and is also radiating.
As a consequence, we expect that certain features of this radiation may carry along characteristic 
imprints of the black hole from where it originated.

This review wants to touch on three basic questions of black hole physics that 
can be addressed by using numerical simulations.
We want to understand
(i) how does the environmental material dynamically behave when it comes close to the black hole,
(ii) how does that material look like for a distant observer, and
(iii) what could be learned about the black hole parameters by comparing theoretical modeling 
and observational data.

In order to answer these questions, the use of numerical simulations is essential, if not 
inevitable.
Fortunately, the availability of large computer clusters for high-performance codes  
has vastly increased over the recent years.
In general, these codes do solve the time-dependent physical equations considering
general relativistic space time.
As a result, the time-evolution of the dynamics of the matter and the electromagnetic 
field and also the radiation field is provided.
Depending on the astrophysical context, different numerical approaches are undertaken.

One option is to concentrate on the dynamics of the material around a black hole.
This is typically the accretion disk that is surrounding the supermassive black holes 
in the center of an active galactic nuclei or the stellar mass black holes in micro quasars.
It is this {\em accretion disk} that makes the black hole {\em shine} \citep{1973A&A....24..337S}.
Depending on the black hole mass and the physical radiation process, we may observe X-ray emission 
or UV light, emission and absorption lines, a Doppler broadening of 
these lines, and we may even use these lines to re-construct the disk structure from the line 
signal (reverberation mapping, \citep{1982ApJ...255..419B}).
The same sources may eject jets - collimated and magnetized beams of high velocity that are 
either launched in the surrounding accretion disk \citep{1982MNRAS.199..883B} or by a 
rotating black hole itself \citep{1977MNRAS.179..433B}.
For both processes a strong magnetic field is essential.
To simulate of the (magneto)hydrodynamics of disks and outflows as well as their radiation
if one of the prime targets of numerical investigations.
What are the mass accretion and ejection rates, the speed and energetics of the 
outflows, or the accretion disk luminosity?

For a black hole located in a intensive radiation field, we may expect to observe its ''shadow"
against the background light as the black hole absorbs all the radiation that enters the
horizon.
For a long time -- that means before the recent detection of gravitational waves -- for many colleagues
only the observation of the black hole shadow would have provided strong evidence for the actual existence 
of black holes.
Numerical simulations of ray-tracing in a strong gravitational field are essential for the understanding 
of a (yet-to-come) observational signal of such a shadow.
The idea is to compare the numerically constructed images of the area close to the 
black hole with the observation of nearby supermassive black hole shadow.
These experiments are currently undertaken and are focusing on the 
black holes with the largest angular diameter that is the one in the Galactic center 
and the one in the active galaxy M87
(see {\tt https://eventhorizontelescope.org/}; {\tt https://blackholecam.org/}).

Even isolated black holes may be observed if the path of light from distant sources is 
penetrating deep enough into the gravitational potential of a black hole that is located 
just between the source and the observer and thus becomes characteristically deflected 
(gravitational lensing; see \citep{1936Sci....84..506E, 1964MNRAS.128..295R}).

Close binary compact objects (neutron stars or black holes) will finally merge and thereby 
strongly distort the space-time around them.
This distortion propagates through space as a gravitational wave which could then be detected on earth.
As this is extremely difficult to measure, only recently such waves could be detected by the LIGO
gravitational wave interferometer (see {\tt https://www.ligo.caltech.edu}).
In order to interpret the observed merger signal, the numerical simulation of a merging binary 
system and the subsequent gravitational wave is a fundamental tool which
can deliver the progenitor masses and even the distance to the gravitational wave source.

%----------------------------------------------------------------------------------------------------------
\section{Numerical approaches}
In this review we will discuss the different numerical methods that were invented to investigate 
the close environment of black holes, considering their increasing numerical complexity.
We distinguish the different approaches in three branches -- these are \\
(i) simulations mainly treating the dynamics of the black hole environment on a fixed metric, \\
(ii) simulations that ray-trace photons that are affected by strong gravity, and \\
(iii) simulations that solve the time-dependent Einstein equations. \\
Of course, a combination of theses approaches are possible, but will be even more demanding.

In the following we first want to briefly compare and summarize the different methods, before we
go on and discuss a few technical details and exemplary results in the remaining sections.

%-------------------------------------------------------------------------------------
\subsection{Dynamics on fixed space-time}
It is considerably the most straight-forward approach to numerically evolve the environment
of a black hole on a fixed space-time, thus applying a fixed metric.
We may understand this approach may as an analogy to the non-relativistic treatment of gravity 
considering a (hydrodynamic) test mass in a Newtonian gravitational potential.
The equations to be solved are the general relativistic (magneto)hydrodynamic (MHD) equations,
that is the equation of motion defined by the time evolution of the energy-momentum-tensor on a 
{\em given} metric, fixed in time.
The general approach is to choose a metric and coordinate system that are convenient for the 
black hole system under consideration and then solve the physical equations of interest.

A typical astrophysical prospect could be to investigate the evolution of an accretion disk or 
torus that is orbiting a black hole.
In particular, one may investigate the stability of disks and tori, or their mass loss by accretion
or the ejection of disk winds or jets.
A topical motivation of this application is to learn about the feedback of supermassive black holes 
for cosmological galaxy formation simulations,
which apply parameterized models of the energy output from galactic black holes.

The approach just described is equivalent to a non-relativistic numerical treatment that
solves the (non-relativistic) MHD equations in the {\em gravitational potential} of e.g. a star.
In both cases, the surrounding masses are {\em test masses} affected by gravity.
As such we cannot tell from pure (magneto-)hydrodynamic simulations their astrophysical masses or
the densities of the objects affected by a fixed gravity.
Typically, the numerical approach gives an accretion rate in {\em code units} with no direct relation
to an astrophysical value.
However, if radiation transport is included (see below) the density scale actually matters
as absorption and emission coefficients depend on the exact physical values 
(see e.g. \citep{2009ApJ...692..411N, 2016ApJ...827...10J}.
Disk luminosities derived from such kind of simulations could be directly compared to observations 
and can thus constrain the disk density and the mass fluxes.

Numerical simulations of the close environment of the black hole were pioneered by Wilson and 
collaborators \citep{1972ApJ...173..431W, 1977mgm..conf..393W, 1984ApJ...277..296H, 1984ApJS...55..211H,2002ApJ...577..866D}.
After some years of silence - probably due to the lack of necessary computer power - the field
of general relativistic time-dependent simulations flourished just after the turn of the
century with a number of groups working on different numerical codes 
\citep{1999ApJ...522..727K, 2001MNRAS.326L..41K, 2002ApJ...577..866D, 2003ApJ...589..458D, 2003ApJ...589..444G, 2006ApJ...641..626N, 2006MNRAS.367.1797M, 2007A&A...473...11D}.

More physics has been added, e.g. radiation processes or a physical resistivity.
Note, however, that for a number of microscopic processes a consistent general relativistic description 
is missing, which may potentially lead to causality issues in the simulation
(see for example \citep{1983AnPhy.151..466H, 2008JPhG...35k5102D} for the case of viscosity or conduction).

An interesting feature to note is that GR-MHD simulations \citep{2003ApJ...589..444G, 2006ApJ...641..626N} 
usually apply (modified) Kerr-Schild coordinates.
In Kerr-Schild coordinates the coordinate singularity at the horizon that is present in the Schwarzschild 
metric or the Kerr metric is avoided by a suitable coordinate transformation.
These codes actually continue to calculate the physical evolution of the variables {\em inside} 
the horizon.
Naturally, physical processes taking place inside the horizon cannot affect the (numerical) world 
outside the horizon. 
The physical conditions defined by the horizon still apply - a neat boundary condition, 
that is difficult to accomplish if the horizon itself is the numerical boundary and if that 
boundary implies physical singularities.

A general difficulty that MHD codes have, are areas of low matter density. 
The numerical time step decreases with increasing Alfv\'en speed and may ''kill" the simulation.
While this is a problem for non-relativistic simulations applying strong magnetic 
fields and low densities, in relativistic simulations the Alfv\'en speed is limited by the speed of
light and thus the time-stepping is limited by the light-crossing time over a grid cell.

Low densities together with a strong magnetic field impose severe problems with certain variable 
transformations that are essential in relativistic MHD codes.
Therefore, relativistic simulations consider a numerical floor model for the critical areas
of low density, keeping the gas density and the internal energy above a ''floor" value, a chosen 
(local) threshold.
By that, the codes can avoid to evolve into an unphysical parameter space, for example negative densities
or pressures.

Typically, the choice of the numerical floor density typically affects the most relativistic regions 
of the simulation, for example a Poynting flux-dominated Blandford-Znajek jet (see below).
For such regions an option may be to apply general relativistic force-free electrodynamic simulations 
\citep{2006MNRAS.367.1797M,2007MNRAS.375..531M}.
However, while such simulation are able to treat highly magnetized areas very well, they hardly can
capture the gas dynamics.

An essential parameter for any simulation using grid codes is the numerical resolution.
High spatial resolution is desirable in order to the resolve the proper physical evolution.
On the other hand, high resolution in a explicit code slows down the simulation because of the 
small numerical time step needed to ensure a causally correct propagation across the grid.
A typical astrophysical context here is the evolution of hydrodynamical instabilities and turbulence, for 
example the magneto-rotational instability (MRI, \citep{1991ApJ...376..214B}).
While the simulations have made great progress towards capturing the MRI and subsequent processes such as 
angular momentum transport or a direct dynamo, the simulations often fail to capture a significant
dynamical range of the turbulent cascade and thus the energy dissipation on small scales.
The is of course a problem that is not restricted to general relativity.

\begin{figure*}[t]
\centering
\includegraphics[width=8cm]{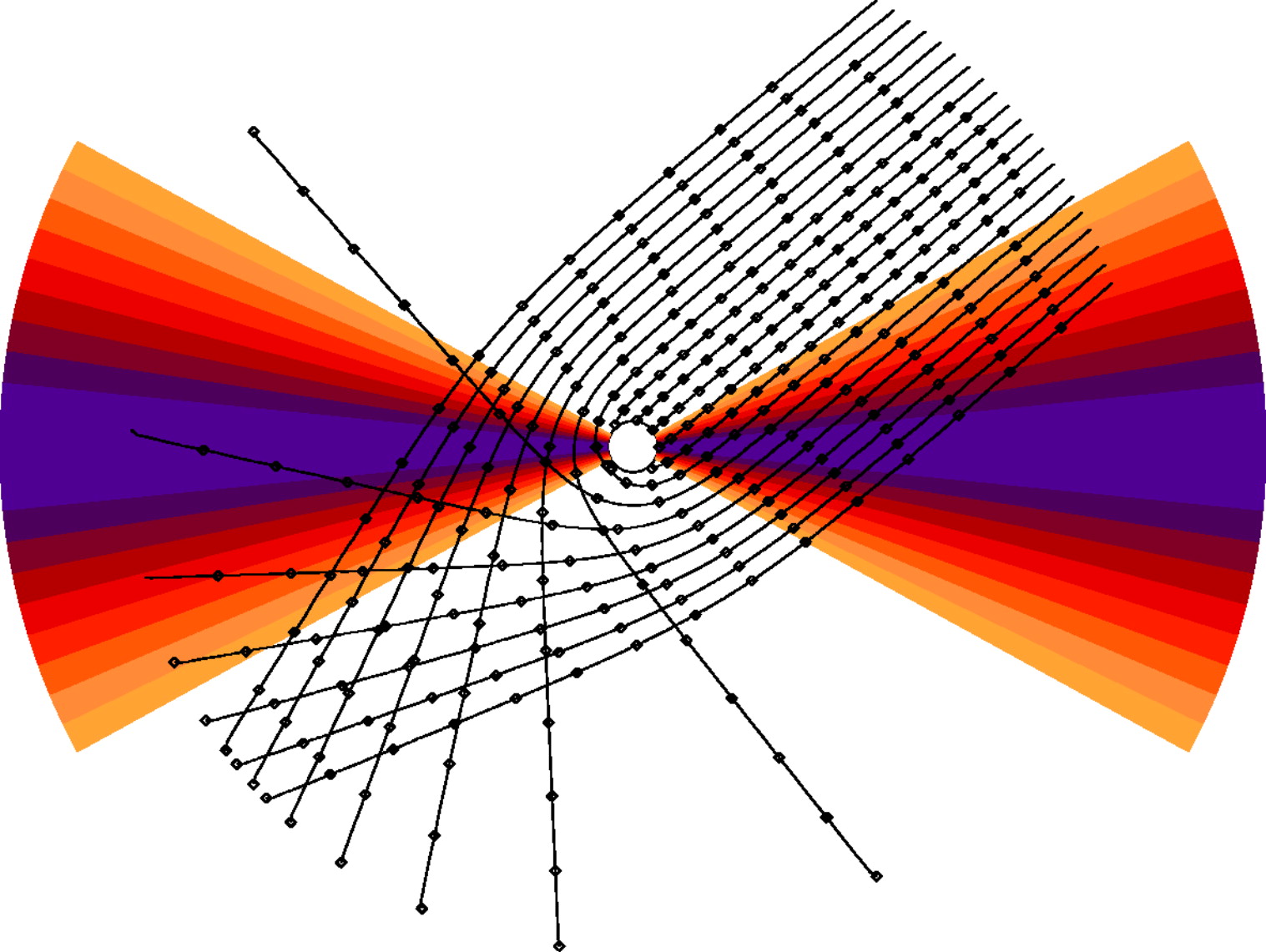}
\caption{Schematic diagram of a typical ray-tracing procedure. The path of a photon 
is integrated along geodesic trajectories from a distant observer to the region 
close to the black hole. 
The photons may either terminate at the horizon (innermost circle) or escape to 
infinity.
The source of photons in this example is an accretion disk with layers
(in color) of different of temperature.
Figure taken from \citep{2006ApJ...651.1031S}.
}
\label{fig:ray-tracing}
\end{figure*}

\begin{figure*}[t]
\centering
\includegraphics[width=5.2cm]{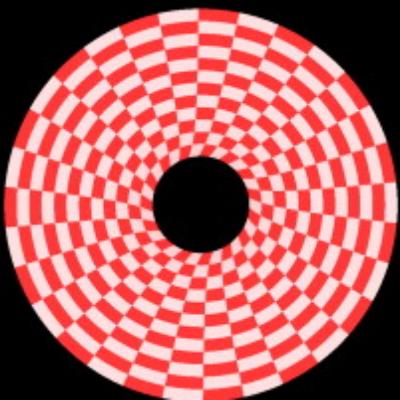}
\includegraphics[width=5.2cm]{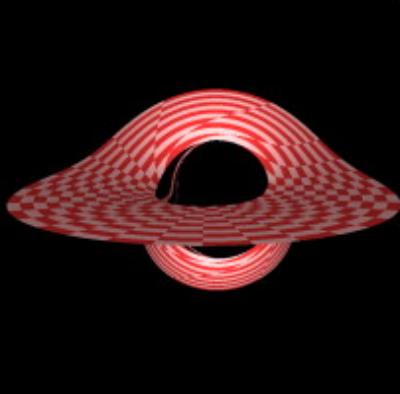}
\includegraphics[width=5.2cm]{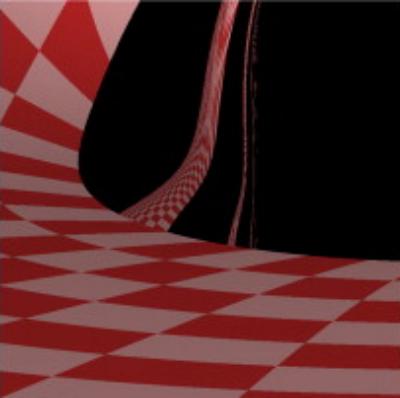}
\includegraphics[width=5.2cm]{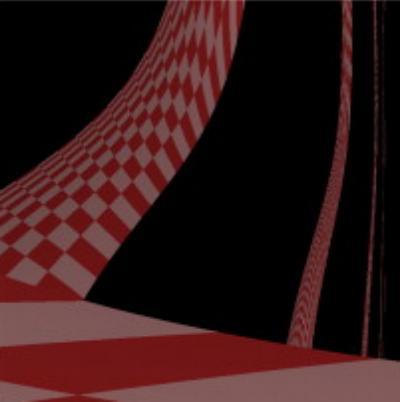}
\caption{
Ray-tracing of a thin accretion disk around an extreme Kerr black hole with mass $M=1$ and
Kerr parameter $a=1$. 
The observer is located at a distance of $50\,M$ at inclination angles of $2\deg$ (left) and $80\deg$ 
(three right panels) with respect to the disk normal. 
The inner and outer disk radii are $r_{\rm in} = 3\,M$ and $r_{\rm out} = 15\,M$.
Higher-order images of the disk become visible within a field of view of $5\deg$ (second 
from right) and $1.5\deg$ (right).
Figure taken from \citep{2014CoPhC.185.2301M}.
}
\label{fig:disk-tracing}
\end{figure*}

%--------------------------------------------------------------------------------
\subsection{Dynamics of interacting compact objects}
Compared to simulations performed on a static metric as discussed above, 
a more general approach is to solve Einstein equations numerically and time-dependent.
This procedure is obviously needed when the mass distribution in the simulation domain
changes substantially in time, leading to a similar change in the metric.
A typical application is the compact object merger.
No fixed metric can be chosen initially - the metric is time-dependent and results from the evolution of the mass distribution.
Further, the change in the metric that is propagated away in space, is just 
known as the gravitational wave.

As for the approach with the fixed metric discussed above, also this approach has non-relativistic equivalent.
The example may be a typical star formation simulation of a molecular cloud collapsing under {\em self-gravity}.
In order to evolve the simulation, after each time step the gravitational potential of the mass distribution has 
to be integrated, before in the next time step the motion of masses is calculated using the updated gravitational 
potential.

Pioneering work on the theoretical framework for numerical general relativity considering e.g. the mass formula for
black holes, a systematic approach to solve Einstein's equations on dynamical space-times, or even a definition of
''numerical relativity" has been published during the 1970s
\citep{1973PhRvL..30...71S,1976PhRvD..14.2443S,1978PhRvD..17.2529S,1979PhRvD..19.2239E}.

It took some years until seminal development of the first 3D numerical codes designed to solve the Einstein equations for 
general vacuum space times succeeded providing the evolution for the first 3D black hole spacetime
\citep{1995PhRvD..52.2059A}.
A major difficulty here that had to overcome is to avoid the singularity inside the horizon or to treat the 
boundary conditions properly for space{\em times}.
Another feature, the so-called apparent horizon becomes important - a kind of spatial boundary that at a certain
time effectively acts as an event horizon.
In a 3D simulations both horizons are time-dependent and may fluctuate as the simulation progresses. 
The difficulty is how to determine the mass inside them?

The problem of dealing with the initial data in case of rotating black holes or distorted non-rotating black holes was treated 
by \citep{1996PhRvD..54.1403B}.
Constructing the initial data for a simulation is a complex task involving to solve Einsteins equations on chosen initial 
spatial coordinates that than has to be embedded in the 4-dimensional space time
(see e.g. \citep{2018PhRvD..98j4011V}).
Another seminal step was to successfully treat the motion of a black hole through a numerical grid.
The difficulty here are the curvature singularities contained within the black hole(s). One way to deal with it is the so-called
black hole-excision by that all parts of teh black hole interior are excluded from the computational domain and only the exterior
region is evolved \citep{1998PhRvL..80.2512C}.

%------------------------------------------------------------------------------
\subsection{Radiation - propagation of light close to a black hole}
The third approach we are discussing is to numerically propagate photons along space-time,
thus to do relativistic radiation transport.
What has discussed above was mainly related to the {\em dynamical} evolution of matter (and the magnetic field).
Simulations of the dynamics deliver the time-dependent distribution of e.g. mass density and velocity.
What can be potentially {\em observed}, is the radiation that is emitted from this material.
Thus, radiation transport will be {\rm the} essential step that allows to compare numerical
simulations of hydrodynamics with the observed radiation pattern.

The path of light that is passing by a black hole is governed by the strong gravity.
The principles seem rather simple:  the trajectory of light rays follows the null geodesics of the metric - the 
Schwarzschild metric for non-rotating black holes or the Kerr metric for rotating black holes.
More complicated situations arise if the metric is time-dependent as for example in case of merging compact 
stars, or if the mass distribution that is emitting and absorbing the light is changing rapidly with time.

Figure~\ref{fig:ray-tracing} shows a schematic diagram of photon paths around a black hole.
The photon trajectories can be bent {\em around} the black hole if a photon comes within the the co-called photon sphere.
Figure~\ref{fig:disk-tracing} displays a 3-dimensional realization of a numerical ray-tracing procedure applied to a 
luminous accretion disk orbiting a Kerr black, as seen by a distant observer under different inclination angles. 
Obviously, a rich radiative structure emerges including parts of the disk that are located behind 
the black hole and become visible by gravity.

The same principle applies and is observationally confirmed also for weaker gravitational fields. 
The famous historical example is the deflection of light by stars predicted by Einstein \citep{1911AnP...340..898E} 
and its confirmation observed during a Solar eclipse \citep{1920RSPTA.220..291D}.
A modern application is the observation of gravitational lensing \citep{1936Sci....84..506E,1964MNRAS.128..295R} 
for a variety of sources. 
Examples are the discovery of planets by micro-lensing events in the Milky Way \citep{2006Natur.439..437B}, 
the observation of multiple images of distant quasars \citep{1985AJ.....90..691H}, 
or the measure of shear and flexion in the large scale structure of the universe (see e.g.~\citep{1992ARA&A..30..311B}).

Apart from the general relativistic effects discussed above, special relativistic Doppler beaming for moving sources 
(a jet or a disk) plays an essential role for the observed signal. 
The blue-shifted signal is Doppler-boosted, thus brighter, while the red-shifted is de-boosted for the observer.

In addition to the pure ray-tracing of light rays trough space-time, radiative transfer can be applied, 
dealing with absorption and emissivity along the path of light.
This problem is substantially more complex as dealing with material opacities or scattering and can be 
solved in approximations.
However, in general, the material properties such as densities, temperatures etc as well as the light sources
can be provided by either a kinematic prescription or by post-processing of dynamical data resulting from an 
(magneto-)hydrodynamic simulation
\citep{2006ApJ...651.1031S, 2010ApJ...717.1092D, 2011CQGra..28v5011V, 2013ApJ...777...13C, 
2015MNRAS.451.1661Z, 2016MNRAS.462..115D, 2016ApJ...820..105P} (see below).

%===================================================================================================
\section{Dynamics \& feedback - (magneto)hydrodynamic simulations}
Black holes are able to feedback to their environment -- implying that mass, energy or angular momentum 
is taken away from the accompanying components or the black hole itself and is deposited further away 
from the black hole.

A typical context is the nuclear activity of galaxy such as quasars with extremely high luminosity that
arises from the black hole accretion disk, or radio galaxies that may drive Mpc-long relativistic jets.
These feedback processes are thought to be essential for cosmological structure formation and the formation of 
galaxies \citep{2012ARA&A..50..455F}.
On a lower energy scales similar processes work in Galactic micro-quasars that are powered by stellar-mass 
black holes \citep{1999ARA&A..37..409M}.
In a similar way, central black holes are believed to be the very origin of the explosive Gamma-ray bursts 
observed on cosmological distances \citep{2015PhR...561....1K}.

A fundamental question that can be tackled by a numerical treatment is the mass accretion to the black hole.
The physics of the accretion process determines the black hole luminosity and also the growth
of the black hole mass (and angular momentum).
Quite a number of general relativistic simulations of black hole accretion flows have been carried out over the 
last decades, both in the hydrodynamic limit 
(e.g. \citep{1972ApJ...173..431W,1984ApJ...277..296H,1984ApJS...55..211H,1991ApJ...381..496H,
2005ApJ...623..347F}),
but also applying MHD, and some of them even considering some sort of radiative transport
(e.g. \citep{2003ApJ...589..444G, 2007ApJ...668..417F,
2009ApJ...692..411N, 2014MNRAS.439..503S,2014ApJ...796..103M}).

The physical origin of magnetized jet outflows has been proposed by seminal papers by \citep{1977MNRAS.179..433B} 
and \citep{1982MNRAS.199..883B}.
The first mechanism - known as Blandford-Znajek mechanism - works only for rotating black holes. 
Here, the frame dragging of space-time leads to the induction of a toroidal magnetic field from the 
poloidal magnetic field threading the black hole ergosphere.
As a result, a strong Poynting flux can be generated, leading to a magnetically dominated jet tower 
that can be filled by lepton pairs that are produced by the strong radiation field of the accretion disk.

The second mechanism - known as Blandford-Payne mechanism - deals with the acceleration and collimation of 
disk winds by magneto-centrifugal forces.
Many non-relativistic MHD simulations on this topic have been published (see e.g. \citep{1997ApJ...482..712O}), 
but also a few relativistic simulations that show that this effect may also work in relativity 
\citep{2010ApJ...709.1100P, 2011ApJ...737...42P}.
However, it seems to be difficult to gain very high Lorentz factors for disk winds \citep{2009MNRAS.394.1182K}.

The launching question - the question how the disk wind material is lifted into the outflow - is more
complex and requires to treat the accretion disk evolution simultaneously.
Analytical (non-relativistic) theory has demonstrated that magnetic diffusivity (respectively, resistivity) is 
essential for launching \citep{1997A&A...319..340F} -- that implies to re-distribute material between different magnetic 
flux surfaces and by that to feed the outflow with disk material.

Simulations of wind or jet launching from thick hot disks have been performed applying non-relativistic codes
\citep{2012ApJ...761..130Y,2015ApJ...804..101Y}, see also \citet{2014ARA&A..52..529Y}.
Similarly, launching simulations of jets from thin disks have been published (see for example \citep{2007A&A...469..811Z,2012ApJ...757...65S,2018ApJ...857...34Z}
pointing again out the importance of the magnetic diffusivity for mass loading.
From non-relativistic simulations we also know that the disk magnetization plays a leading role for governing the
outflow properties \citep{2016ApJ...825...14S}.

General relativistic MHD simulations
were pioneered by \citep{1977mgm..conf..393W}, while first results were published by
\citep{1999ApJ...522..727K}.
Soon, further numerical codes became available boosting the number of publications in this field
\citep{2002ApJ...577..866D, 2003ApJ...589..444G, 2003ApJ...589..458D,
2003ApJ...592.1060D, 2004ApJ...611..977M, 2005ApJ...620..878D, 2006ApJ...641..626N, 
2006MNRAS.367.1797M, 2009ApJ...692..411N, 2012MNRAS.426.3241N,2012MNRAS.426.3241N,2017ComAC...4....1P}.
Over the last decade constant and rapid progress in this field has been made.

Most often for the simulation a stationary hydrodynamic torus \citep{1976ApJ...207..962F} is assumed as initial condition 
from which a (thin) disk may emerge.
However, it looks like that some leading parameters of the flow evolution seem to be determined by this 
(probably not realistic) initial setup and are not "forgotten" during the simulation, so its use
is maybe limited (see e.g. \citep{2013A&A...559A.116P}).
The torus-disk structure evolves under the magneto-rotational instability \citep{1991ApJ...376..214B}, leading to 
accretion of material towards the black hole and also to weak disk winds. 
Advection of magnetic flux along the disk towards the black hole leads to the accumulation of a strong axial poloidal flux 
that ''threads" the black hole.

The feasibility of the so-called Blandford-Znajek mechanism was numerically demonstrated by
\citep{2001MNRAS.326L..41K} by a pure electromagnetic simulations and later also 
by a number of MHD simulations in general relativity 
(see for example \citep{2003PhRvD..67j4010K, 2003ApJ...589..444G, 2004ApJ...611..977M, 
                      2005ApJ...620..878D, 2005ApJ...630L...5M, 2005MNRAS.359..801K}).

\begin{figure}
\centering
\includegraphics[width=8.5cm]{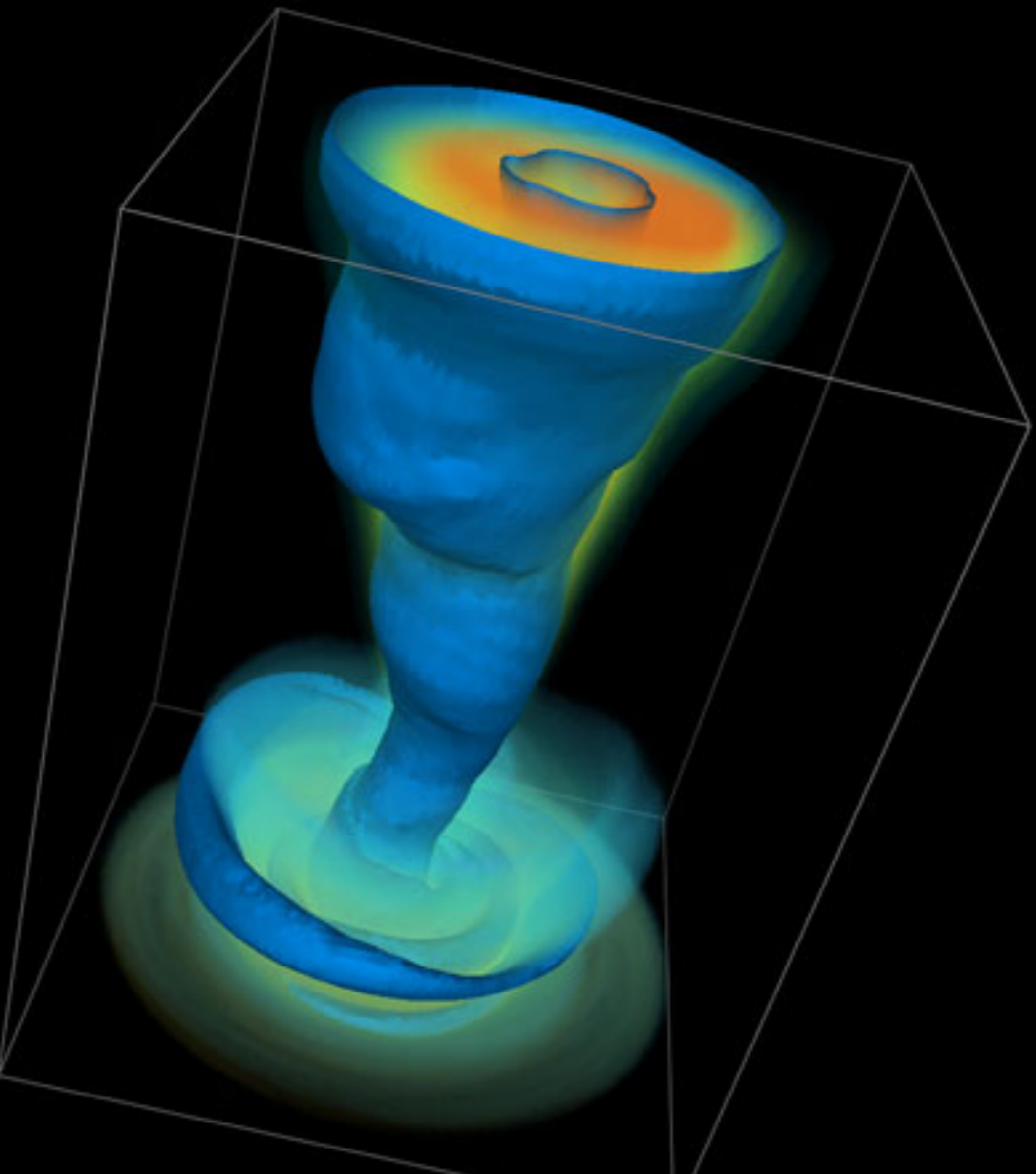}

\caption{Jet launching simulation applying 3D GR-MHD.
Shown is volume rendering of the internal energy density (log scale) for the relativistic black hole jet
within a box of $350\times350\times1000$ black hole gravitational radii $R_{\rm g}$ after $t=4000$ time units
$t_{\rm g} = R_{\rm g} / c$.
Only the jet, not the counter jet is shown, and fragments of the perturbed disk wind.
The jet reaches Lorentz factors of $\Gamma$ \lax 10 and a half-opening angle of $\Theta_{\rm jet} \simeq 5^{\circ}$.
Figure taken from \citep{2009MNRAS.394L.126M}.
}
\label{fig:mckinney2009}
\end{figure}

Figure~\ref{fig:mckinney2009} shows the result from a 3D GR-MHD simulation investigating the stability of 
jets launched by a rotating black hole \citep{2009MNRAS.394L.126M}. 
The authors find jets with Lorentz factor of about 10 emerging along the black hole rotational axis with opening angles of 
about 10 degrees. 
The jet reaches about 1000 gravitational radii distance from the black hole without significant distortion by 
a 3D instability.
These numbers are constrained by the numerical capability of the available codes.
So far much higher relativistic jets cannot be modeled, meaning that the jet mass fluxes and the jet speed are limited by the 
floor model (see above).
The jet stability is essentially affected by the numerical resolution and also by the conditions in the environment.

%-----------------------------------------------

Of particular interest is the so-called plunging region between the inner disk radius (located approximately 
at the innermost stable orbit, ISCO) and the horizon.
Of course, also the generation of jets driven by black hole rotation is governed by this area.
This region is in particular sensitive to the frame dragging of a spinning black hole and also determines the energy 
conversion from the potential and orbital energy of the disk material. 
Examples for simulations emphasizing on the plunging region are \citep{2010MNRAS.408..752P} who 
investigate how the electromagnetic stresses within the ISCO depend on the black hole spin 
or the thickness of the accretion disk;
or \citep{2011MNRAS.418L..79T} who consider jet formation in from magnetically arrested disks.
In \citep{2010ApJ...711...50T} the interrelation between black hole spin and the jet power output 
applying their results to the long-standing problem of the radio loud/quiet dichotomy of AGN.
In a similar approach \citep{2012MNRAS.423L..55T} find that black holes that rotate prograde compared 
to the disk may trap more magnetic flux then retrograde black holes leading to a higher jet efficiency.

The time-evolution of thin disks has been studied by \citep{2010ApJ...711..959N} in particular considering 
the evolution of and the electromagnetic stress at the inner disk radius in the Schwarzschild case. 
This is important in order to understand the efficiency of the accretion process and the
time scale for black hole growth.
Tilted thin disks and the potential onset of precession due to the Bardeen-Petersen effect
were investigated by 
\citep{2014ApJ...796..103M} in 3D simulations lasting up to 13,000 time units 
$t_{\rm g} = R_{\rm g} / c$
that are measured in light crossing time over the gravitational radius $R_{\rm g}$.
Precession of the central black hole or the inner accretion disk may be seen as a
time-dependent alignment of the jet outflow on much larger (pc-kpc) scale.

Also resistive GR-MHD have been applied, in particular to investigate the magnetic field generation 
by a dynamo process \citep{2013MNRAS.428...71B, 2014MNRAS.440L..41},
to compare the energy output from the disk wind and the Blandford-Znajek jet 
\citep{2017ApJ...834...29Q, 2018ApJ...859...28Q}, 
or a stellar magnetized collapse to a black hole
\citep{2009MNRAS.394.1727P, 2013PhRvD..88d4020D}.

%---------------------------------------------------------------------------------------
\section{Radiation from simulated gas dynamics}
From GR-MHD simulations of the black hole environment the time evolution of the gas dynamics and 
the magnetic field can be derived. 
However, what is observed in reality is the radiation that is emitted from this 
gas - observers obtain spectra or radiation images.
In order to compare simulation results with observations, it is therefore essential to couple
radiation transfer to the hydrodynamics of the simulation.

Most of the emission from the black hole source is expected to come from the innermost accretion disk.
However, the light that is emitted in the plunging region during the final infall may actually carry 
most of potential (general) relativistic imprints, in particular imprints from the metric that is
defined by the central black hole.
By simulating these signals and comparing them to observations we may gain direct information about 
the black hole characteristics such as e.g. the black hole spin
(see e.g. \citep{2012MNRAS.424.2504Z}).

Combining MHD dynamics and radiation transfer is computationally highly expensive and yet numerically 
and physically still limited. 
A number of issues have to be considered.
The method of choice would be to do radiation transfer modeling along with the hydrodynamics
that is to follow the absorption and emission of photons by the gas along with the gas dynamics.

It is well known that the radiation transfer equations need to be {\em closed}. 
A closure relation is needed as radiation transfer is typically treated considering the {\em moments} such as
radiation energy, radiation flux and radiation pressure. 
The local geometry of the radiation field needs to be specified by a radiation model (for example LTE)
which closes the set of radiation transfer equations.
A few astrophysical options for the closure problem have been established.
Each method may have its advantages and disadvantages for a certain astrophysical problem.
One option is the flux-limited diffusion, applied for a short mean free path of photons, thus for large
optical depth close the the case of LTE.
The application of the diffusion approximation is often coarse, however, radiative transfer would be very expensive.
A well known closure is the Eddington approximation, typically applied in a plane-parallel geometry like a geometrically 
thin stellar photosphere. However, this is probably not a good choice for a turbulent disk-jet structure around a black hole.

The method of choice applied in most GR-MHD simulations is the so-called M1 closure that evolves the radiation flux 
independently from the radiation energy. 
The basic assumption is that the radiation field is locally a dipole aligned with the radiation flux.
The M1 closure has been introduced in GR hydrodynamics by \citep{2013MNRAS.429.3533S} and for special relativistic 
magnetized fluids by \citep{2013ApJ...772..127T}.
In contrast to the diffusion limit, M1 could handle "shadows".
In \citep{2014MNRAS.439..503S} this approach was applied to investigate the impact of opacity on the accretion 
flow structure and evolution in case of super-Eddington luminosities.
In a similar way, \citep{2014MNRAS.441.3177M} could model super-Eddington disks with accretion rates 100-200 times the 
Eddington accretion rate.

Another closure that has recently be proposed is the Monte-Carlo closure for general relativistic radiation 
hydrodynamics \citep{2018MNRAS.475.4186F}.
This has been applied in particular also to investigate the neutrino transport.
Neutrino transport has been proven to be essential for the understanding the neutrino emission they radiate along with 
the gravitational waves when forming a black hole \citep{2016PhRvD..93d4019F}.

A computationally less expensive approach that avoids the costly radiative transfer modeling is to consider a
cooling function in the GR-MHD simulation.
This method has been introduced in GR-MHD by \citep{2009ApJ...692..411N} to calculate the radiative efficiency of the 
black hole disk. The disk radiation has been ray-traced to compute the flux that eventually is received by the observer.
In \citep{2007CQGra..24S.259N} the mm and sub-mm flux for Sgr A* has been derived. 
Spectra and radiation maps were derived applying various combinations of the black hole spin parameter and the disk accretion rate
in order to match the observed 1-mm flux of 4 Jy. 
In \citep{2010ApJ...717.1092D} the authors compare the numerically derived disk luminosity to observations of Sgr\,A* deriving
an accretion rate of $5^{+15}_{-2}\times 10^{-9}\msun\,{\rm yr}^{-1}$ as well as temperatures and inclination angle.
Similarly, \citep{2015ApJ...799....1C} derive spectra from GR-MHD modeling and compare them to the properties of Sgr\,A*.
In particular, they show that by combining this information with the emission images certain models (those with strong 
funnel emission) can be ruled out.
Heating and cooling of electrons have been included in recent GR-MHD simulations by \citep{2015MNRAS.454.1848R}
and \citep{2017ApJ...844L..24R}.

A widely used approach to construct black hole radiation features is post-processing of the simulated MHD data.
This is an option for example when considering the {\em non-thermal radiation} or optically thin cases.
For non-thermal emission, however, the energy distribution of the emitting high energy particles cannot be 
derived self-consistently within the MHD approach, and must thus be prescribed.
Depending on whether the particle energies are interrelated with the gas pressure, the gas density or the
magnetic field energy, the resulting radiation maps may vary greatly (see e.g. \citep{2011ApJ...737...42P}).

Post-processing of radiation features is always problematic when dealing with optically thick media.
The optical depth of the material and also its opacity is not always known and post-processing remains 
an approximation.
This is in particular an issue when dealing with {\em thermal emission}. 
Then, ideally, radiation transport should be done along with the MHD simulation (as discussed above), as the 
radiation field may affect the gas temperature that in turn governs the radiation field.

We mention another issue for this method that is linked to the coordinate system and time stepping generally
used by GR-MHD codes.
GR-MHD simulations are typically performed in (modified) Kerr-Schild coordinates, thus providing time 
slices - snapshots - in Kerr-Schild time.
Observations - ray tracing - is typically done in the observers frame, for example in Boyer-Lindquist 
coordinates.
The difficulty is that in every snapshot in Kerr-Schild coordinates - representing a 2D or 3D distribution 
of e.g. gas density or magnetic field - each pixel represents a {\em different time} in Boyer-Lindquist 
coordinates.
Thus, performing a ray-tracing along the photon path that is smooth in time seems impossible.

In addition to that, ray-tracing across material that moving relativistically needs to consider the 
{\em retarded times}. 
The problem is that while the signal (emitted at a certain retarded time) is propagating with time 
through the gas, ''new" material is moving {\em into} that path of the photon (this motion is
actually) calculated by the GR-MHD simulation.
It is therefore be difficult to interpret post-processed radiation maps for highly time-variable sources 
such as jets or accretion disk coronae.

So far, ray tracing in GR-MHD-simulated gas dynamics circumvents this problem by assuming the so-called 
{\em fast-light approximation}
that postulates that the light travels faster than the dynamics of the system evolves.
Another option to get (post-processed) mock images of simulation data is to stack simulation snapshots and 
ray-trace along the time-averaged gas distribution.
While this is a valid approach in order to obtain time-averaged radiation features such as a typical spectrum of the
source, the method is obviously not applicable for highly time-variable phenomena.

In reality, post-processing of radiation transfer may not be fully consistent with the dynamics of the relativistic plasma, 
but at the moment, there seem to be no numerical tools at hand that can circumvent this problem in respect of the 
available CPU resources.
Post-processing radiative transfer is widely applied
\citep{2006ApJ...651.1031S, 2009ApJS..184..387D, 2010ApJ...717.1092D, 2013ApJ...777...13C, 2015MNRAS.451.1661Z, 2016MNRAS.462..115D,
2016ApJ...831....4P,2016ApJ...820..105P,2017ApJ...845..160P}
and has recently been extended also for comptonization \citep{2016MNRAS.457..608N}.

\begin{figure*}
\centering
\includegraphics[width=16cm]{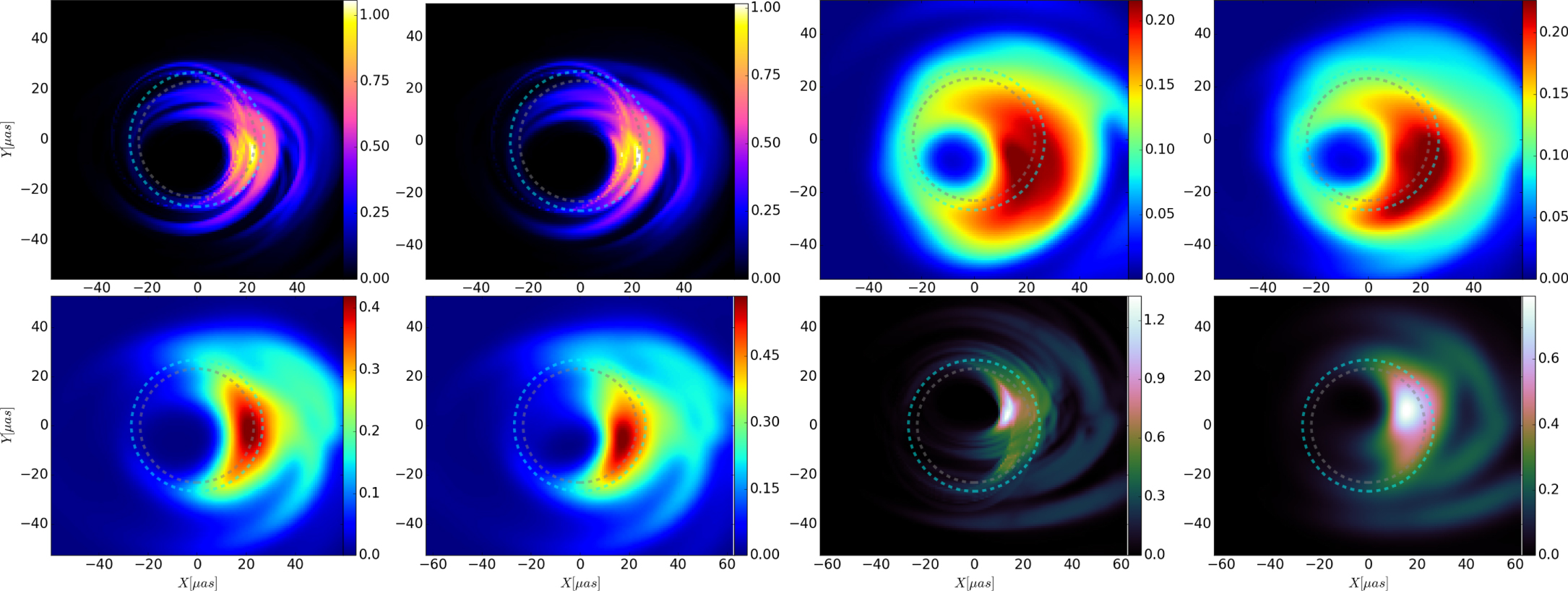}
\caption{Black hole shadow image plane intensity plots for the Galactic center black hole Sgr~A*
as calculated from a model geometry consisting of a disk and jet around a black hole.
The series demonstrates the richness in the parameter space that can be considered, such as
(i) a single observer comparison to the parameter space of \citep{2010ApJ...717.1092D};
(ii) instead eight snapshots time-averaged;
(iii) applying an inclination of  $i=45^\circ $, and a fixed ion-to-electron temperature 
(as in Moscibrodzka et al. (2009); 
(iv) applying another default time instead (little difference); 
(v) applying different jet and gas temperatures (no isothermal jet); 
(vi) applying an isothermal jet; 
(vii) a best-fit for all parameters; and 
(vii) also including scattering. 
Figure taken from \citep{2017ApJ...837..180G}.
}
\label{fig:gold2017}
\end{figure*}

As a specific example, we mention \citep{2006ApJ...651.1031S} who derive light curves of black hole accretion disks from 
radiative transfer based on GR-MHD simulations with the application on quasi-periodic oscillations of 
X-ray binaries.
In \citep{2017ApJ...837..180G} a comparison study of black hole shadow images for Sgr~A* is presented
demonstrating the impact of e.g. the inclination angle and the temperature 
prescription, scattering, or the stacking of images on the visibility of the shadow.
This is shown in Figure~\ref{fig:gold2017} demonstrating the complexity of the parameter space and thus the difficulty 
to obtain ''true" radiation maps.
The authors conclude that all these effects together {\textit may make it difficult to unambiguously detect the black hole shadow 
or to extract information about the space-time from its size and shape, before the plasma physics and dynamical 
properties of disk and jet are constrained} (see also \citep{2015ApJ...814..115P}).

Nevertheless, mock observations from simulated MHD simulations of the black hole environment are extremely important for 
understanding the observations of the "shadow" of nearby supermassive black holes (Sgr\,A$^*$ and M87)
as proposed by the Event Horizon Telescope (EHT) initiative (see {\tt https://eventhorizontelescope.org/}.
Radiation transfer modeling is essential in order to disentangle the various processes at work at these spatial scales,
such as disk and outflow dynamics, light bending, or absorption or scattering \citep{2015ApJ...799....1C, 2017ApJ...845..160P}.

%-----------------------------------------------

%--------------------------------------------------------------------------------
\section{Compact object mergers}
{\em ''For a brief moment, a binary black hole merger can be the most powerful astrophysical event
in the visible Universe"} \citep{2017PhRvD..96b4006K}.
The numerical treatment of compact object mergers is considerably the most difficult numerical task to date.
While the dynamical evolution of the material around a single black hole can be treated on a fixed space-time,
the evolution of (close) binary black holes is substantially more complex.
As the metric changes with time, for this problem the Einstein equations needs to be solved numerically 
in a way that 
(i) allows for a well-posed initial-boundary-value problem, and that 
(ii) singularites inherent in the black hole spacetimes can be circumvented
(see e.g.~\citep{2010RvMP...82.3069C, 2015CQGra..32l4011S}).
The same challenges appear for mergers of binary neutron stars into a black hole.

In comparison to the simulations discussed in the last sections, most simulations of black hole mergers typically consider only gravity.
Therefore, the results depend only on few parameters (black hole masses and spins), and do not rely on further assumptions
on physical quantities as e.g. the accretion disk density or the magnetic field.
Fundamental breakthrough in the simulation of binary black holes came around 2005 with seminal papers considering the three basic 
phases of a black hole merger - the orbit (only up to a single orbit in these early years), the merger, and the following ringdown
\citep{2004PhRvL..92u1101B, 2005PhRvL..95l1101P, 2006PhRvL..96k1102B, 2006PhRvL..96k1101C}.

\begin{figure*}
\centering
\includegraphics[width=7.5cm]{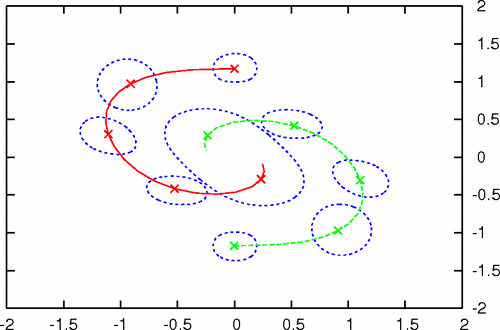}
\includegraphics[width=7.5cm]{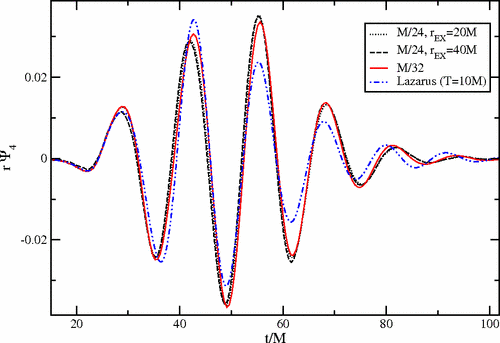}

\caption{Early example of a merging binary black hole simulation.
Positions of the apparent horizons (left) with time, $t=0, 5, 10, 15, {\rm and} 20M$. 
The line shows the trajectories of centroids of the individual apparent horizons (dashed).
Gravitational wave signal (right) for simulation runs of different resolution.
Figures taken from \citep{2006PhRvL..96k1102B}.
}
\label{fig:baker2006}
\end{figure*}

Deriving the gravitational wave signal from a merger in its final stage requires to evolve the 3D Einstein equations.
Further, the merger simulation has to run sufficiently long \citep{2001PhRvL..87A1103A, 2006PhRvL..96l1101D}.
An early example of the gravitational wave signal and the apparent horizons of the merging black hole as derived 
by \citep{2006PhRvL..96k1102B} is shown in Figure~\ref{fig:baker2006}.
These early modeling efforts succeeded with using excision to exclude singular regions within the horizons (see above).
For the extraction of the wave signal from the numerical simulation of the merger -- thus to extract the 
radiative part of the numerical solution of the Einstein equations -- a number of methods have been developed. 
For a detailed discussion of the problems involved we refer to \citep{2016LRR....19....2B}.

\begin{figure}
\centering
\includegraphics[width=8.9cm]{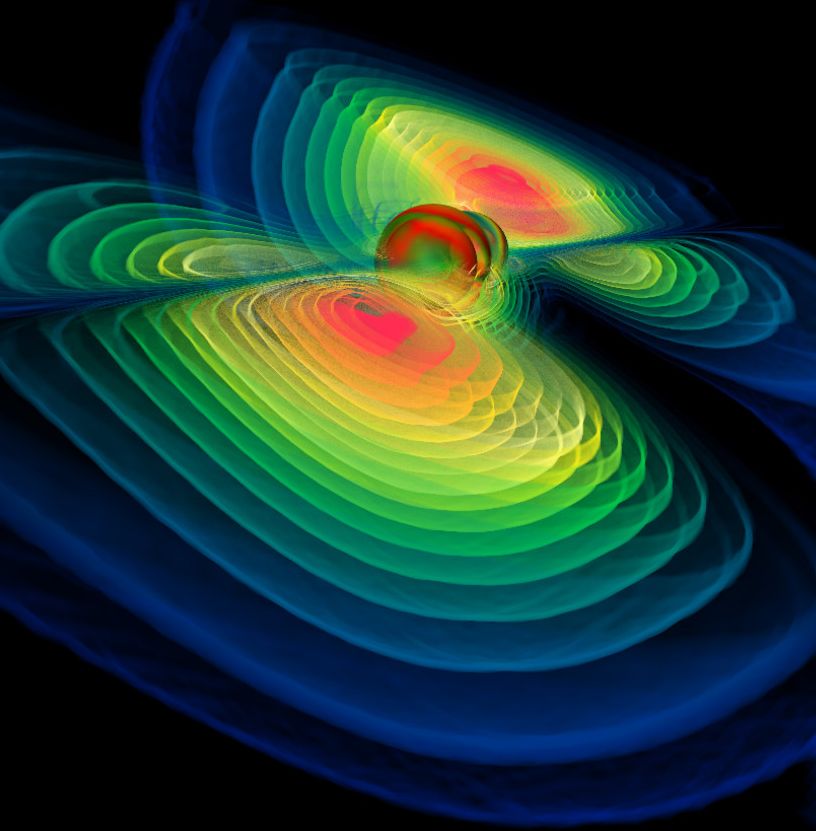}
\caption{
A numerical simulation showing the gravitational radiation emitted by the violent merger of two black holes.
Figure taken from \citep{Bruegmann2014}.
}
\label{fig:bruegmann2014}
\end{figure}

The field of compact merger simulations has evolved rapidly, benefiting also from the advance in CPU power.
Some codes such as the ''Einstein Toolkit" have even been made publicly available \citep{2012CQGra..29k5001L,2014CQGra..31a5005M}.
Hundreds of binary black hole simulations have been performed by now, allowing to do a proper statistics of gravitational 
wave forms and the gravitational wave luminosity or on the final black hole properties based on different initial mass 
ratios or black hole spins (see e.g. \citep{2016CQGra..33t4001J, 2017CQGra..34v4001H, 2017PhRvD..96b4006K}.
A recent, colorful 3D representation of the merging black holes emitting gravitational waves is shown in
Figure~\ref{fig:bruegmann2014}.

As considerably one of the most fascinating events in recent science, the long lasting efforts in both theoretical 
predictions and detector experiments were culminating in 2015 when LIGO interferometer eventually detected
a gravitational wave signal \citep{2016PhRvL.116f1102A, 2016PhRvL.116x1102A} that was perfectly matching the 
theoretically predicted waveform models for a black hole merger \citep{2014PhRvL.113o1101H, 2014PhRvD..89f1502T} .
The detection clearly indicates on a black hole merger of masses of $36^{+5}_{-4} \msun$ and $29^{+4}_{-4} \msun$ 
to a final black hole of mass $62^{+4}_{-4}\msun$ at a luminosity distance of $410^{+160}_{-180}$ Mpc, 
thereby radiating away gravitational waves\footnote{
Interested readers can actually redo the modeling of the GW~150914 binary black hole merger gravitational wave signal
by using the ''Einstein Toolkit", see {\it https://einsteintoolkit.org/gallery/bbh/}.}
of energy of $3.0^{+0.5}_{-0.5} \msun c^2$.
Besides the fascinating confirmation of the existence of gravitational waves, the data also revealed a new and unexpected mass 
range for black holes. 
As of early 2019, gravitational wave detections has reached the number of 11 
(for the actual status see {\it https://www.ligo.caltech.edu/page/detection-companion-papers)}.

%-----------------------------------------------
%-----------------------------------------------

Another class of merger simulations consider mergers of neutron star binaries.
This field has been pioneered by Shibata \& Uryu \citep{2000PhRvD..61f4001S,2003PhRvD..68h4020S, 2005PhRvD..71h4021S}.
Binary neutron star mergers are related to black hole numerics as these mergers {\em do form} a black hole.
As a matter of fact, these kind of mergers may provide an observational signal in addition to the pure gravitational wave.

A particular difficulty in neutron star merger simulations compared to black hole merger simulations is 
that one has to consider{\em  material} with a certain equation of state.
However, despite a huge observational and theoretical progress, the equation of state for neutron stars is
still one of the major unsolved problems in astrophysics (see e.g. \citep{2013ApJ...765L...5S,2016ARA&A..54..401O}).

Merger simulations investigating different nuclear equations of state have been performed \citep{2011PhRvD..83l4008H}, 
indicating for example that different kinds of final merger states can be realized that are strongly dependent
on the equation of state applied. 
Simulations of this kind also allow to tackle the evolution of a massive torus surrounding the remnant black hole.
In turn, the gravitational wave signal from binary neutron star mergers could be used to probe 
the extreme-density matter in these stars \citep{2017ApJ...842L..10R}.

Examples for further specifics of neutron star mergers in comparison to black hole merges are tidal effects of the stellar 
mass distribution \citep{2015PhRvL.115i1101B}, 
a possible mass ejection and radiation \citep{2016PhRvD..93l4046S}, 
or neutrino emission \citep{2016PhRvD..93d4019F}.
Also, neutron star mergers have long been discussed as sources of short gamma-ray bursts, therefore simulations 
of such a scenario are essential to confirm this hypothesis \citep{2015PhRvD..91l4056B, 2017CQGra..34h4002P}.
A particularly exciting example of this kind of simulations is the modeling \citep{2017PhRvD..96l3012S} 
of the gravitational wave signal GW170817 that was connected to the simultaneous detection of a gamma ray burst
\citep{2017PhRvL.119p1101A}.

A remaining step to do would be the combination of the two approaches discussed above, namely the combination of a merger 
simulation with the magnetohydrodynamics of the binary black hole environment,
thus to combine the numerical approach discussed in this section (pure gravity) with the dynamical 
MHD simulations discussed before.
This has been achieved by e.g.~\citep{2012PhRvL.109v1102F}, who, for the first time, run a fully general relativistic MHD 
simulations of an equal-mass black-hole binary merger in a magnetized, circum-(black hole)-binary accretion disk 
(see also \citep{2014PhRvD..89f4060G}).
Such simulations can predict for example the change of accretion rate during the merger event.
Essentially, they may also provide - in addition to the binary merger gravitational wave signal - the contemporary 
electromagnetic signal during the merger event.

%-----------------------------------------------

%---------------------------------------------------------------
\section{Summary \& outlook}
This brief (and therefore incomplete) review intended to summarize the rapid development and
the various applications of numerical methods in general relativity over the past decades.

The numerical approaches involving black holes can roughly be distinguished in simulations
(i) treating the dynamics of the black hole environment on a fixed metric; or 
(ii) solving for time-dependent solutions of the Einstein equations that allow to to follow 
the merger of compact objects into a remnant black hole and the gravitational waves that are 
emitted; or
(iii) following the photons that are affected by strong gravity, and thus allowing the see how 
black holes ''look alike".
Even more demanding are combinations of these approaches such as obtaining mock observations
of simulated dynamical data involving magnetohydrodynamics and radiation transfer.

The comfortable situation today is that quite a number of codes have been developed for these different 
approaches.
In addition, also the computational resources to operate them have vastly improved. 
An essential point is to compare the existing codes and the robustness of their results
concerning e.g. the accretion towards the black hole.
This task is currently undertaken as a collaboration of nine groups and first results are
published recently \citep{Porth2019}.

Besides fundamental questions of theoretical physics such as e.g. testing general relativity 
in the strong field limit,
from the astrophysical perspective the most intriguing problems are maybe
(i) the formation and early growth of (distant) supermassive black holes - involving most 
probably both, the merger scenario and also disk accretion;
(ii) the mystery of gamma-ray bursts, or
(iii) the inner structure (the equation of state) and the evolution of compact stars.

Essential observational input is expected from future gravitational wave signals measured by LIGO,
as well as from highly resolved radio observations that are going to image the shadow of 
nearby supermassive black holes.
Striking first images of a black hole shadow namely that of the nearby galaxy M87 
and corresponding simulations in GR-MHD that are essential for the interpretation of the data
have been published by the EHT collaboration very recently \cite{2019ApJ...875L...1E,2019ApJ...875L...4E}.

With all the advance in the observational data that will become available, 
the numerical modeling will be central for the understanding of these findings.

%%%%%%%%%%%%%%%%%%%%%%%%%%%%%%%%%%%%%%%%%%
%\acknowledgments{In this section you can acknowledge any support given which is not covered by the author 
%contribution or funding sections. This may include administrative and technical support, or donations in kind (e.g., materials used for experiments).}

%%%%%%%%%%%%%%%%%%%%%%%%%%%%%%%%%%%%%%%%%%
\conflictsofinterest{The author declares no conflict of interest.} 

%%%%%%%%%%%%%%%%%%%%%%%%%%%%%%%%%%%%%%%%%%
%% optional
\abbreviations{The following abbreviations are used in this manuscript:\\

\noindent 
\begin{tabular}{@{}ll}
3D & three-dimensional \\
AGN & Active Galactic Nuclei\\
EHT & Event Horizon Telescope \\
GR-MHD & General relativistic magnetohydrodynamics\\
ISCO & Innermost stable circular orbit \\
LIGO & Laser Interferometer Gravitational-Wave Observatory\\
MHD & Magnetohydrodynamics\\
\end{tabular}
}


%=====================================
% References, variant A: internal bibliography
%=====================================
\reftitle{References}

% The following MDPI journals use author-date citation: Arts, Econometrics, Economies, Genealogy, Humanities, IJFS, JRFM, Laws, Religions, Risks, Social Sciences. For those journals, please follow the formatting guidelines on http://www.mdpi.com/authors/references
% To cite two works by the same author: \citeauthor{ref-journal-1a} (\citeyear{ref-journal-1a}, \citeyear{ref-journal-1b}). This produces: Whittaker (1967, 1975)
% To cite two works by the same author with specific pages: \citeauthor{ref-journal-3a} (\citeyear{ref-journal-3a}, p. 328; \citeyear{ref-journal-3b}, p.475). This produces: Wong (1999, p. 328; 2000, p. 475)

%% for journal Sci
%\reviewreports{\\
%Reviewer 1 comments and authors’ response\\
%Reviewer 2 comments and authors’ response\\
%Reviewer 3 comments and authors’ response
%}

%%%%%%%%%%%%%%%%%%%%%%%%%%%%%%%%%%%%%%%%%%
\end{document}